\documentclass{elsart}

\begin{document}


\def\Bid{{\mathchoice {\rm {1\mskip-4.5mu l}} {\rm
{1\mskip-4.5mu l}} {\rm {1\mskip-3.8mu l}} {\rm {1\mskip-4.3mu l}}}}


\runauthor{Mark S. Byrd and Paul B. Slater}

\begin{frontmatter}
\title{Bures Measures over the Spaces of 
Two and Three-Dimensional Density
Matrices}

\author[UT]{Mark S. Byrd\thanksref{SMU}}
\author[ISBER]{Paul B. Slater}

\address[UT]{Physics Department,
The University of Texas at Austin,
Austin, TX 78712-1081}

\address[ISBER]{ISBER,
University of California,
Santa Barbara, CA 93106-2150}

\thanks[SMU]{Visiting the Department of Physics, 
Southern Methodist University, Dallas, TX 75275-0175, 
email: mbyrd@physics.utexas.edu}



\begin{abstract}
Due to  considerable recent interest in the use of density matrices for 
a wide variety of purposes, including quantum 
computation, we present a general method for their
parameterizations in terms of Euler angles.  
We assert  that this is of  more 
fundamental importance than (as several people have 
remarked to us) ``just another parameterization of the 
density matrix.''  There are several uses to which this methodology
can be put. One that has received particular attention
is in the construction of certain distinguished 
(Bures) measures  on the $(n^2 -1)$-dimensional
convex sets of $n \times n$  density matrices.
\end{abstract}

\begin{keyword} 
Density matrix; Euler angles; Bures metric; Haar measure
\end{keyword}

\end{frontmatter}




\section{Introduction}

The density matrix \cite{blum} --- having  origins in 
early (independent)
work of Landau and von Neumann --- has proved to be a 
very useful concept in physics.  Researchers have 
devoted substantial efforts in 
describing the spaces defined by density matrices
\cite{bloore}, in using them to analyze the separability 
of quantum systems \cite{zhsl,slatjpa}, in
comparing information-theoretic properties of various probability 
distributions over them 
\cite{slatpla}, as well as studying the question of parallel 
transport in this context \cite{hubner}.  There is, of course, a quite
 straightforward manner in which to 
parameterize 
density matrices --- 
simply in terms of the real and complex parts of the entries of
these Hermitian matrices.
A more indirect
approach, but one of particular interest, relies upon
Euler angle parameterizations of the special unitary matrices.  
It has been known for quite some 
time that one can parameterize density matrices 
with the help of such angles.  
To do so, we take a diagonal density matrix, 
$\rho$, which 
represents our quantum system in a particular basis.  We then perform 
a unitary transformation in Hilbert space that takes $\rho$ 
to an arbitrary basis. This unitary transformation can be 
assumed  to have a determinant 
equal to one since an overall phase does not affect 
the physics.  Now when we apply  the 
unitary transformation, we operate in the  
following manner,
$$
\rho^\prime = U \rho U^\dagger,
$$
where $U\in SU(n)$ for an $n$-state system.  The diagonal matrix $\rho$ 
 can 
be parameterized by its $n-1$ eigenvalues, and the unitary transformation by 
$n^2-1$ 
variables.  At this point,
however,  we have parameterized ${\rho}'$ using 
{\sl too many}, that is $n^2 +n -2$ parameters, since the density matrices
for the $n$-state systems comprise 
only an $(n^2-1)$-dimensional convex set.
A primary objective here will be to demonstrate how such an
``over-parameterization'' can be avoided, by the elimination 
(from the unitary transformation) of 
$n-1$ of the set of $n^2 +n -2$ parameters.

Below, we will show that  the density matrices for the 2-state and 3-state 
systems can be conveniently and insightfully 
 parameterized in terms of Euler angle 
coordinates. In doing so, we  take full advantage of 
the group properties associated with the unitary transformation.  In 
particular, we obtain  useful (Bures) 
measures on the spaces of density matrices, 
which can be used in the
integration of functions over spaces and subspaces of quantum systems.  
Specifically, in Section \ref{twostatepar} we exhibit the Euler angle-based  
parameterization of the 2-state 
quantum systems.  In Section \ref{threestatepar} 
we present the Euler angle-based 
parameterization of the 3-state systems.  In Section 
\ref{nstatepar} we discuss their straightforward generalization 
to $n$-state systems.  
In Section \ref{twostatemeasure} we show how to use this to 
write down the Bures measures
 on the space of density matrices for 2-state 
systems and in Section \ref{threestatemeasure} its counterpart 
for 3-state systems.  Finally, we discuss ongoing work which is intended
to yield 
the explicit generalization to  4-state systems. This, we anticipate 
will be highly useful in analyzing the 
(fifteen-dimensional) space occupied by pairs of qubits 
(cf. \cite{slatexact}).

\section{Parametrization of Density Matrices}


\subsection{2-state Density Matrices}
\label{twostatepar}

In this section we review certain properties of density matrices 
for the 2-state systems and demonstrate how to parameterize the 2-state 
density matrix in Euler angles.  This should make the extension 
to 3-state systems in the following section 
more transparent.

Much is known about 2-state density matrices.  
One of the more obvious properties 
is that a 2-state system can be spanned by two pure 
state density matrices of the following form:
\begin{equation}
  \rho_1 = \left(\begin{array}{cc}
           1 & 0 \\
           0 & 0 
           \end{array}\right),\;\;\;\;
  \rho_2 = \left(\begin{array}{cc}
           0 & 0 \\
           0 & 1
           \end{array}\right).
\end{equation}
(These can be denoted by antipodal points on a two-sphere, making use of 
the 
elegant Riemann sphere representation \cite{penrose}.) For a mixed 
state diagonal 
density matrix, we write the linear combination of these two 
density matrices as
\begin{equation}
\rho = \sum \rho_i a^i,
\end{equation} 
where $\sum a^i = 1$.  The $a^i$ may be represented in several ways.  
One common way is to let $a^1 = a$ and $a^2 = 1-a$.  What we wish to 
emphasize in this paper is that they are better represented by 
$a^1 = \cos^2 \theta$ and $a^2 = \sin^2 \theta$.  Why this is true will be
more obvious in the next section.

To take $\rho$ to an arbitrary configuration (basis) 
we act with a unitary transformation $U \in SU(2)$, in the manner 
\begin{equation}
\rho \rightarrow \rho^\prime = U \rho U^{-1} = U \rho U^\dagger.
\end{equation}
One may note that an $n$-state density matrix should have $n^2 - 1$ 
parameters, whereas this one appears to have 4, that 
is $3 \;(\mbox{from}\; SU(2))
+1\;(\mbox{from the diagonal})\;$.  However, in the Euler angle 
parameterization of $SU(2)$, given by 
\begin{equation}
U = e^{i\sigma_3 \alpha}e^{i\sigma_2 \beta}e^{i\sigma_3 \gamma},
\end{equation}
where the $\sigma$s are the Pauli matrices, we see that the parameter 
$\gamma$ drops out since $\sigma_3$ is diagonal 
and so the matrix 
exponential $e^{i\sigma_3 \gamma}$ commutes with $\rho$ 
leaving precisely three parameters, as is appropriate and desired.  

An important part of this parameterization is the ranges of the 
angles.  For the 2-state case these are well-known (up to normalization 
given in {\it eg.}, \cite{lb}).
They are
\begin{equation}
0 \leq \alpha \leq \pi, \;\;\; 0 \leq \beta \leq \pi/2,\;\;\; 
0 \leq \theta \leq \pi/4.
\end{equation}


\subsection{3-state Density Matrices}
\label{threestatepar}

For 3-state density matrices we can achieve a similar parameterization.  
The pure states are spanned by
\begin{equation}
\rho_1 = \left( \begin{array}{lcr}
                     1 & 0 & 0 \\
                     0 & 0 & 0 \\
                     0 & 0 & 0 
                \end{array}\right),\;\;\;
\rho_2 = \left( \begin{array}{lcrr}
                     0 & 0 & 0 \\
                     0 & 1 & 0 \\
                     0 & 0 & 0 
                \end{array}\right),\;\;\;
\rho_3 = \left( \begin{array}{lcr}
                     0 & 0 & 0 \\
                     0 & 0 & 0 \\
                     0 & 0 & 1 
                \end{array}\right).
\end{equation}
We can then obtain a mixed state by the linear combination $\sum \rho_i a^i$, 
with $\sum a^i = 1$.  Here again we use the squared components of a 
 sphere.  
In this case, it is a two-sphere, thus for a 3-state system we take a 
generic mixed state to be
\begin{equation}
\rho = \left( \begin{array}{ccc}
   \cos^2 \theta_1 \sin^2 \theta_2 &      0                    &    0       \\
             0             & \sin^2 \theta_1 \sin^2 \theta_2 &    0       \\
             0             &      0                    & \cos^2 \theta_2 
              \end{array} \right).
\end{equation}

We then can  take this to an arbitrary configuration (basis) by
the process
\begin{equation}
\rho \rightarrow \rho^\prime = U \rho U^\dagger,
\end{equation}
where $U \in SU(3)$.  In the Euler angle parameterization of $SU(3)$, $U$ is 
given by \cite{eapar}
\begin{equation} \label{SU(3)}
U = e^{i\lambda_3 \alpha} e^{i\lambda_2 \beta} e^{i\lambda_3 \gamma}
 e^{i\lambda_5 \theta} e^{i\lambda_3 a} e^{i\lambda_2 b}
 e^{i\lambda_3 c} e^{i\lambda_8 \phi/\sqrt{3}}
\end{equation}
where the Gell-Mann matrices have been used;
\begin{equation}
\begin{array}{crcr}
\lambda_1 = \left( \begin{array}{crcl}
                     0 & 1 & 0 \\
                     1 & 0 & 0 \\
                     0 & 0 & 0   \end{array} \right), &
\lambda_2 = \left( \begin{array}{crcr} 
                     0 & -i & 0 \\
                     i &  0 & 0 \\
                     0 &  0 & 0   \end{array} \right), &
\lambda_3 =  \left( \begin{array}{crcr} 
                     1 &  0 & 0 \\
                     0 & -1 & 0 \\
                     0 &  0 & 0   \end{array} \right), \\
\lambda_4 =  \left( \begin{array}{clcr} 
                     0 & 0 & 1 \\
                     0 & 0 & 0 \\
                     1 & 0 & 0   \end{array} \right), &
\lambda_5 = \left( \begin{array}{crcr} 
                     0 & 0 & -i \\
                     0 & 0 & 0 \\
                     i & 0 & 0   \end{array} \right), &
 \lambda_6 = \left( \begin{array}{crcr} 
                     0 & 0 & 0 \\
                     0 & 0 & 1 \\
                     0 & 1 & 0   \end{array} \right), \\
\lambda_7 = \left( \begin{array}{crcr} 
                     0 & 0 & 0 \\
                     0 & 0 & -i \\
                     0 & i & 0   \end{array} \right), &
\lambda_8 = \frac{1}{\sqrt{3}}\left( \begin{array}{crcr} 
                     1 & 0 & 0 \\
                     0 & 1 & 0 \\
                     0 & 0 & -2   \end{array} \right).
\end{array}    
\end{equation}
Since $\lambda_3$, and $\lambda_8$ appear on the right 
in (\ref{SU(3)}), they commute with 
the diagonal
matrix $\rho$  and drop out of the parameterization.  Thus we are left with 
 $3^2-1 = 8$ parameters for the 3-state case, as we should expect 
\cite{bloore}.  

As in the 2-state case, the ranges of the angles in the parameterization 
are very important.  For the case of three states we have
\begin{equation}
0 \leq \alpha,\gamma,a \leq \pi, \;\;\;\; 
0 \leq \beta, \theta, b \leq \pi/2, 
\end{equation}
\begin{equation}
   0 \leq \theta_1 \leq \pi/4, \;\mbox{and}\; 0 \leq \theta_2 \leq \cos^{-1}(1/\sqrt{3}). 
\end{equation}

Here one should note that using $a,\;b$ and $1-a-b$ for the diagonal 
elements of the density matrix would give rise to a domain of integration 
that is non-rectangular.  For density matrices of higher dimensional 
systems this could be extremely akward.  Here however, we see that 
the domain is indeed a rectangular solid with the final angle for a 
density matrix of $n$-dimensions having the range 
$0\leq \theta_{n-1} \leq \cos^{-1}(1/\sqrt{n})$.


\subsection{$n$-state Density Matrices}
\label{nstatepar}

Though there is now no explicit parameterization of $SU(n)$ in 
terms of Euler angles, the generalization is rather obvious.  The 
difficulty in manipulating the one-forms to calculate the volume 
elements increases as $n^2$ so unless some substantial 
progress is made in getting computers to manipulate trigonometric 
functions symbolically, it appears that progress will 
be quite slow.  It is anticipated that the density-matrix-parameterization
program advanced here will not be achieveable 
for groups much higher in dimensionality than $dim(SU(4)) = 15$.  
Perhaps, however, with sufficient interest and effort it could be accomplished
 for the Lie group $SU(8)$, 
in order to effectively describe the 
63-dimensional space of {\it three} entangled qubits.


\section{Bures Measures on the
Convex Sets of Density Matrices}


\subsection{2-state Systems}
\label{twostatemeasure}

The parameterization 
of the space of density matrices by Euler angles immediately 
leads to a particularly simple procedure for integrating
 over the space.  A natural (Bures) measure on the space is 
given by the product of the (Hall) measure 
\cite[eqs. (24, 25)]{hall} \cite{slathall}  on the space of 
eigenvalues ($\lambda_{1},\ldots,\lambda_{n}$), that
is 
\begin{equation} \label{hv}
\mbox{d}u = 
{\mbox{d} \lambda_{1} \ldots \mbox{d} \lambda_{n} \over (\lambda_{1} 
\ldots \lambda_{n})^{1 \over 2}} \prod_{j < k} 4 {(\lambda_{j} -\lambda_{k})^2
\over \lambda_{j} + \lambda_{k}}
\end{equation}
 and  
the measure on the 
the space of unitary matrices, the (modified/truncated) Haar measure.  
The geometry of 
the 
three-dimensional space is obvious from the parameterization.  We have a solid section 
of the two-sphere and the space $SU(2)/U(1) \cong S^2$.  In other words, our 
measure is given by 
\begin{equation}
DV = du\times dS^2,
\end{equation}
where $DV$ is the 
Bures  measure on the space of density matrices, $dS^2$ is the standard 
measure on $S^2$, {\it ie.} $dS^2 = d(G/H)$, with $G=SU(2)$ and $H=U(1)$ 
\cite{BrockerandDieck} and $du $ is given by (\ref{hv}) with $n=2$.


\subsection{3-state Systems}
\label{threestatemeasure}

The measure on the 
eight-dimensional space of density matrices describing 3-states is 
a generalization of the previous case.  We have 
\begin{equation}
DV = du\times d(G/H),
\end{equation}
where $DV$ is the Bures measure on the space of density matrices, 
with $G=SU(3)$ and $H=U(1)\times U(1)$ 
\cite{BrockerandDieck} and $du$ is given by (\ref{hv}) with 
$n=3$.  The ranges of the Euler angles 
are given in Section \ref{threestatepar}.  This is a new result.  
In the past the ranges of the angles were not properly specified 
when using the Haar measure. Additionally, 
the problem of overparameterization \cite{zhsl}
was not clarified until recently \cite{us}.

In \cite{slathall}, it has been shown how the Bures measures for the
2- and 3-state systems can be normalized to form probability distributions.

\section{Applications/Conclusions} 

The point of this note has been to clarify that: 
(1)  the Euler angle parameterization eliminates
any  naive over-parameterization (\cite{zhsl}) of 
the density matrix using a unitary group representation; and 
(2)  doing so helps in providing a natural (Bures) 
measure on the space of density matrices, which incorporates 
the (truncated) Haar measure on the group manifold of $SU(n)$.

So this is not ``just another parameterization of the density matrix'', 
while
proving very useful for many calculations.  We find that people often do 
not appreciate the point of view of Feynman, who wrote
 that ``every theoretical 
physicist who is any good knows six or seven different theoretical 
representations for exactly the same physics'' \cite{Feyman}.

In addition this gives rise to a rather natural 
(Bures) measure on the space of 
density matrices, using which one can integrate 
functions of quantum-mechanical interest (such as the
von Neumann entropy, $-\mbox{Tr} \rho \log{\rho}$) over spaces and subspaces 
of the group with relative ease.  Much of the geometry of the space 
can then be seen in terms of the moduli of the group space \cite{us}, 
a point apparently not appreciated by a number of readers of \cite{us}.


\section{Acknowledgements}

M.S.B. would like to thank the DOE for partial support of this work 
under grant number DOE-ER-40757-123 and Southern Methodist University 
for the hospitality of the Group and for the use of the facilities 
funded by DE-FG03-95E40908.  P.B.S. is grateful to the Institute for
Theoretical Physics for computational support.



\end{document}